\documentclass[twocolumn,aps,prc,superscriptaddress]{revtex4}

\usepackage{amssymb}
\usepackage{amsmath}
\usepackage{graphicx}
\usepackage[normalem]{ulem}
\usepackage{multirow}
\usepackage{appendix}
\usepackage{CJK}
\usepackage[usenames]{color}
\usepackage{bm}
\begin{document}
\title{Simulating chiral anomalies with spin dynamics}
\author{Wen-Hao Zhou}
\affiliation{Shanghai Institute of Applied Physics, Chinese Academy of Sciences, Shanghai 201800, China}
\affiliation{University of Chinese Academy of Sciences, Beijing 100049, China}
\author{Jun Xu\footnote{corresponding author: xujun@zjlab.org.cn}}
\affiliation{Shanghai Advanced Research Institute, Chinese Academy of Sciences, Shanghai 201210, China}
\affiliation{Shanghai Institute of Applied Physics, Chinese Academy of Sciences, Shanghai 201800, China}

\begin{abstract}
Considering that the chiral kinetic equations of motion (CEOM) can be derived from the spin kinetic equations of motion (SEOM) for massless particles with approximations, we simulate the chiral anomalies by using the latter in a box system with the periodic boundary condition under a uniform external magnetic field. We found that the chiral magnetic effect is weaker while the damping of the chiral magnetic wave is stronger from the SEOM compared with that from the CEOM. In addition, effects induced by chiral anomalies from the SEOM are less sensitive to the decay of the magnetic field than from the CEOM due to the spin relaxation process.
\end{abstract}

\maketitle


The spin dynamics can induce many interesting phenomena in various systems with different types of spin-orbit couplings~\cite{Mor15}. For instance, in nucleonic systems where nucleons are massive compared with nuclear spin-related interactions, the nuclear spin-orbit coupling is responsible for the magic number in finite nuclei~\cite{May49,May55} as well as various spin-related observables in heavy-ion reactions~\cite{Xu15}. In the latter case, the spin dynamics can be described by the spin-dependent equations of motion derived from the spin-dependent Boltzmann-Vlasov equation~\cite{Xia16}. In the massless limit, the Dirac equation is decoupled, and particles with different chiralities are affected by the Weyl spin-orbit coupling, which leads to chiral anomalies and various of interesting phenomena in not only material science (see, e.g., Ref.~\cite{RMP18}) but also heavy-ion physics~\cite{Kha16,Hua16}.

Due to the chiral symmetry restoration of partons at high energy densities reached in relativistic heavy-ion collisions, the above massless limit is approximately satisfied. This may cause the chiral magnetic effect (CME)~\cite{Alex80,Kha08,Fuku08} induced by the non-zero axial charge density, leading to the net electric charge current along the direction of the magnetic field, and the resulting electric charge separation can be measured indirectly by comparing the correlation between particles of the same and opposite electric charges~\cite{Vol04,STAR1,STAR2,ALICE1}, though there are recently hot debates on the background contribution~\cite{Back1,Back2,Back3,Back4,Back5}. A dual effect is called the chiral separation effect (CSE)~\cite{CSE1,CSE2} induced by non-zero electric charge density, leading to the net axial charge current along the direction of the magnetic field. The interplay between the CME and the CSE can produce a collective excitation called the chiral magnetic wave (CMW)~\cite{CMW1,Kha111,CMW2}, which may generate an electric quadrupole moment in a quark matter, and can be responsible for the elliptic flow splitting between particles of opposite electric charges~\cite{Kha112,STAR3,ALICE2}.

The various phenomena mentioned above originated from the Weyl spin-orbit coupling can be studied with transport simulations based on the chiral kinetic equations of motion (CEOM) derived from different approaches~\cite{CKM1,CKM2}. It is interesting to see that the CEOM can also be derived from the spin kinetic equations of motion (SEOM) for massless particles by using an $\hbar$ expansion of the spin with respect to the momentum in the adiabatic limit~\cite{Zhou}. However, such expansion limits the validity of the CEOM only for particles with larger momenta~\cite{CKM1}. As shown in Ref.~\cite{Zhou}, an artificial truncation is needed in the momentum space in transport simulations, and this may underestimate the effects induced by chiral dynamics compared with the theoretical limits. It is also a question how important are the higher-order $\hbar$ terms and how good is the adiabatic approximation in deriving the CEOM from the SEOM. In the present study, we do transport simulations directly according to the SEOM, and compare the results of the CME and CMW with those from the CEOM.


We start from the Hamiltonian for massless spin$-1/2$ particles with the charge number $q=\pm 1$ and the helicity $c=\pm 1$ under an external magnetic field $\vec{B}$
\begin{equation}\label{h}
H=c\vec{\sigma} \cdot \vec{k},
\end{equation}
where $\vec{\sigma}$ are the Pauli matrices, $\vec{k}=\vec{p}-e\vec{A}$ is the kinematic momentum with $\vec{p}$ being the canonical momentum and $\vec{A}$ being the vector potential of the magnetic field $\vec{B}$. Considering $\vec{\sigma}$ as the expectation direction of the spin in the classical limit, the SEOM from Eq.~(\ref{h}) can be written as
\begin{align}
        \dot{\vec{r}}      &= c\vec{\sigma},                              \label{equ:spin1} \\
        \dot{\vec{k}}      &= c\vec{\sigma} \times qe\vec{B},             \label{equ:spin2} \\
        \dot{\vec{\sigma}} &= c\frac{2}{\hbar}\vec{k} \times \vec{\sigma}.\label{equ:spin3}
\end{align}
Equation~(\ref{equ:spin1}) shows that the velocity is always the speed of light, Eq.~(\ref{equ:spin2}) indicates the Lorentz force, and Eq.~(\ref{equ:spin3}) describes the precessional motion of $\vec{\sigma}$ with respect to $\vec{k}$. It is easy to prove that the above SEOM conserve the single-particle energy $c\vec{\sigma} \cdot \vec{k}$. Using the following approximation~\cite{Huang,huangy}
\begin{equation}\label{app}
\vec{\sigma} \approx c\hat{k} - \frac{\hbar}{2k}\left( \hat{k} \times \dot{\hat{k}} \right),
\end{equation}
one gets the CEOM~\cite{CKM1,CKM2}
    \begin{align}
        \sqrt{G}\dot{\vec{r}} &= \hat{k} + \hbar \left( c\vec{b} \cdot \hat{k} \right) qe\vec{B}, \label{equ:chi1} \\
        \sqrt{G}\dot{\vec{k}} &= \hat{k} \times qe\vec{B},                                        \label{equ:chi2}
    \end{align}
with $\sqrt{G}=1+\hbar( qe\vec{B} \cdot c\vec{b})$, where $\hat k=\vec{k}/k$ is a unit vector and $\vec{b}\!=\!\vec{k}/\!\left(2k^3\right)$ is the Berry curvature for massless particles. Equation~(\ref{app}) quantifies the small deviation between $\vec{\sigma}$ and $c\hat{k}$ in the $\hbar$ order, and assumes that $\vec{\sigma}$ evolves much faster with time than $\vec{r}$ and $\vec{k}$. With Eq.~(\ref{app}), the single-particle energy becomes $c\vec{\sigma}\cdot\vec{k}=k$ for the CEOM. Considering $\vec{r}$, $\vec{k}$, and $\vec{\sigma}$ as free variables, the SEOM do not change the phase-space volume~\cite{Rei80}. On the other hand, Eq.~(\ref{equ:chi2}) in the CEOM modifies the phase-space integral $\mathrm{d}^3r\mathrm{d}^3k/\left(2\pi\hbar\right)^3$ to $\sqrt{G}\mathrm{d}^3r\mathrm{d}^3k/\left(2\pi\hbar\right)^3$~\cite{Xiao05}, so the average value of any statistical quantity is correspondingly calculated according to $\langle A \rangle = \sum_i A_i \sqrt{G_i} / \sum_i  \sqrt{G_i}$ by taking $\sqrt{G_i}$ for the $i$th particle as a weight factor in transport simulations.


    The simulation was studied in a cubic box system with the periodic boundary condition under a uniform external magnetic field, with the side length of the cubic box $2l = \mathrm{10\,fm}$. Two scenarios of simulations have been done, i.e., by using the SEOM [Eqs.~(\ref{equ:spin1})-(\ref{equ:spin3})] and using the CEOM [Eqs.~(\ref{equ:chi1})-(\ref{equ:chi2})]. The initial momentum distribution is sampled according to $1/\left\{ \exp[(k-\mu_{qc})/T]+1 \right\}$, where $\mu_{qc}\!=\!q\mu + c\mu_{5}$ is the chemical potential with $\mu(\mu_5)$ being the electric (axial) charge chemical potential, and $T$ is the temperature of the system set as 300 MeV. The density of each particle species is determined by the temperature and the corresponding chemical potentials, and all particles are uniformly distributed in the coordinate space. For the SEOM scenario, the initial particle spins are set as $\vec\sigma=c\hat{k}$. For the CEOM sceinario, the momentum distribution is expected to be unchanged. For the SEOM scenario, the equilibrated momentum and spin distributions are expected to be anisotropic in the direction along and perpendicular to the magnetic field, but the analytical expression is not available. However, since the energy $E=c\vec{\sigma}\cdot \vec{k}$ for each particle is conserved, $\mu_{qc}$ and $T$ are the same chemical potential and temperature at the equilibrated state for the SEOM scenario, with the energy distribution $1/\left\{ \exp[(E-\mu_{qc})/T]+1 \right\}$ unchanged during the evolution. The dynamics include not only the EOM but also two-body scatterings, with the isotropic scattering cross section $\sigma_{22}$ determined by the specific shear viscosity of the value ${\hbar}/{4\pi}$ (see the appendix of Ref.~\cite{Zhou} for details). The collision probability for a pair of particles with the energy $E_1$ and $E_2$ in a volume $(\Delta x)^3$ and a time interval $\Delta t$ is~\cite{Zhe05}
    \begin{equation}
        P_{22} = v_{rel}\sigma_{22}\frac{\Delta t}{(\Delta x)^3},
    \end{equation}
    where $v_{rel}=s/\!\left(2E_1E_2\right)$ is the M$\phi$ller velocity with $s$ being the square of the invariant mass of the particle pair. In the simulation, only particle pairs in the same cell with the volume $(\Delta x)^3=1\, \mathrm{fm^3}$ can collide with each other, and the time step is set as $\Delta t=0.01\,\mathrm{fm/c}$. The particle momentum after each scattering is sampled isotropically in the center-of-mass frame of the collision pair. For the SEOM scenario, the expectation direction of the particle spin $\vec{\sigma}^\prime$ after each scattering is determined by the energy conservation condition $c\vec{\sigma} \cdot \vec{k} = c\vec{\sigma}^\prime \cdot \vec{k}^\prime$ and the maximum polarization condition, i.e., assuming that the spin always tends to be polarized in the magnetic field so that $qe\vec{B}\cdot\vec{\sigma}^\prime$ should have the maximum value. Whether an attempted scattering can become a successful one is decided by the Pauli blocking probability $1-(1-f_1^\prime)(1-f_2^\prime)$. The occupation probability $f_{1/2}^\prime$ is $1/\left\{ \exp[(k-\mu_{qc})/T]+1 \right\}$ for the CEOM scenario, while for the SEOM scenario it is calculated by counting the phase-space occupation probability for particles with different charge numbers and spin states with respect to the magnetic field. There is no truncation needed for the SEOM scenario, and an artificial truncation $0.3<\sqrt{G}<1.7$ is used for the CEOM scenario to remove particles with too small $k$~\cite{Zhou}.

    The CME and the CSE are generally expressed by the following relations
    \begin{align}
        \vec{J}   &= \alpha\frac{N_c}{2\pi^2\hbar^2}\mu_5e\vec{B}, \label{equ:J}\\
        \vec{J}_5 &= \alpha\frac{N_c}{2\pi^2\hbar^2}\mu e\vec{B},   \label{equ:J5}
    \end{align}
     where $N_c=3$ is the color degeneracy, $\vec{J}$ and $\vec{J}_5$ are the electric and axial current, and $\alpha=1$ corresponds to the theoretical limit. In simulations based on the CEOM, the truncation to remove small-$k$ particles generally underestimates the CME and the CSE~\cite{Zhou}, leading to $\alpha<1$. To understand the CME and the CSE from the SEOM, we take the second-order derivative of $\dot {\vec{r}}$ with respect to $t$, i.e.,
     \begin{equation}
     \frac{d^2 \dot {\vec{r}}}{dt^2} = \frac{2}{\hbar}cq(\vec{\sigma}\times e\vec{B})\times \vec{\sigma} + c\frac{4}{\hbar^2} \vec{k}\times(\vec{k}\times\vec{\sigma}),\notag
     \end{equation}
     where particles with different $c$ and $q$ become separated. We will see that the linear relations between the currents and the chemical potentials are also approximately satisfied for the SEOM scenario, but both the CME and the CSE are weaker, leading to $\alpha<1$. Under an external magnetic field in the $+y$ direction, the equation describing the CMW as an interplay between the CME and the CSE can be written as~\cite{CMW1,Kha111}
    \begin{equation}
        \left(\partial_t \pm v_p \partial_y- D_L\partial_y^2\right) \rho_{R\!/\!L} = 0, \label{equ:cmw}
    \end{equation}
    where
    \begin{equation}
        v_p =\alpha \frac{3\hbar eB}{2\pi^2T^2} \label{equ:v_p}
    \end{equation}
    is the phase velocity related to the magnitude of the CME and the CSE, $D_L$ is the longitudinal diffusion constant characterizing the damping of the CMW. We take the convention that the upper (lower) sign is for right-handed ($qc>0$) particles (left-handed ($qc<0$) particles) with the number density $\rho_R$ ($\rho_L$).


    \begin{figure}[ht]
        \includegraphics[angle=0,scale=0.31]{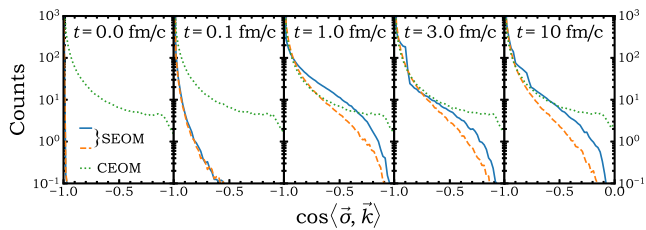}
        \caption{(Color online) Time evolution of the cosine angle distribution between $\vec{k}$ and $\vec{\sigma}$ from the SEOM without the collision process (solid), the SEOM with the collision process (dashed), and the CEOM with the collision process (dotted) under the magnetic field $eB_y=0.5\,\mathrm{GeV\!/fm}$ in the $+y$ direction.}
        \label{fig:angle_s_k}
    \end{figure}

We first look at the cosine angle distribution between $\vec{k}$ and $\vec{\sigma}$, which is defined as $\arccos[(\vec{\sigma}\cdot\vec{k})/k]$ for the SEOM scenario and $\arccos[(c\dot{\vec{r}}\cdot\vec{k})/|\dot{\vec{r}}|k]$ for the CEOM scenario, in Fig.~\ref{fig:angle_s_k}. Only half of the distribution within $(-1,0)$ is displayed, since it is symmetric within $(-1,1)$. Due to the initialization for the SEOM scenario, $\cos\langle \vec{\sigma},\vec{k} \rangle$ is $\pm 1$ in the initial stage, while the distribution evolves gradually away from $\pm 1$ and becomes equilibrated in the final stage. It is seen that without the collision process, the time evolution of the cosine angle distribution has some oscillation behavior towards equilibrium, compared with the scenario with the collision process where the oscillation is largely damped. For the CEOM scenario, the cosine angle distribution doesn't change with time, and is broader compared with that from the SEOM. The strength of effects induced by chiral anomalies is closely related to the distribution of the angle between $\vec{k}$ and $\vec{\sigma}$.

    \begin{figure}[ht]
        \includegraphics[angle=0,scale=0.53]{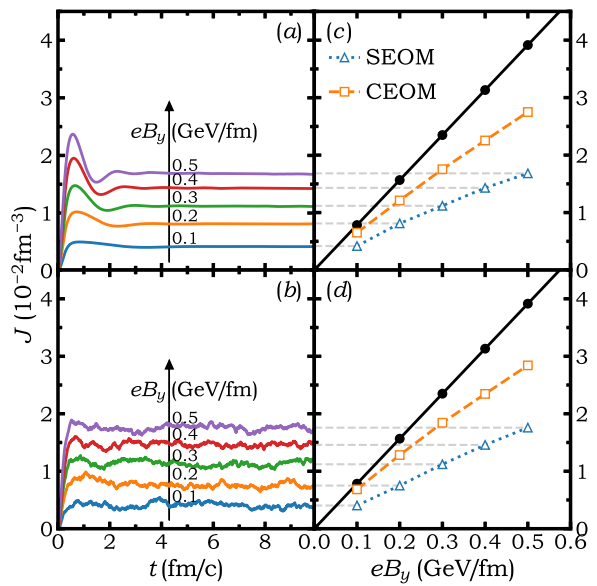}
        \caption{(Color online) Left: Time evolution of the electric charge current with the initial axial chemical potential $\mu_5=20\,\mathrm{MeV}$ under different strengths of the magnetic field from the SEOM scenario without (a) and with (b) the collision process; Right: Dependence of the equilibrated electric charge current on the strength of the magnetic field from the SEOM, as well as those from the CEOM and the theoretical limit (solid line) plotted for comparison. }
        \label{fig:no_sca}
    \end{figure}

    The left panels of Fig.~\ref{fig:no_sca} display the time evolution of the electric currents $J$ generated by the magnetic field of different strengths and an axial chemical potential $\mu_5=20\,\mathrm{MeV}$ for the SEOM scenario. The electric current is calculated as $J=n\langle q (\dot{\vec{r}})_y \rangle$, with $n$ being the total number density of all particles and $\langle q (\dot{\vec{r}})_y \rangle$ being the average charge velocity in $y$ direction. Since the momentum and the spin are sampled according to an isotropic distribution, $J$ is zero in the initial stage. As the time evolves, the electric charge current is induced. A stronger oscillation behavior is observed for the SEOM scenario without the collision process as shown in panel (a) compared with that with the collision process as shown in panel (b), consistent with that observed in Fig.~\ref{fig:angle_s_k}. Although the oscillation is stronger under a stronger magnetic field, the equilibrated values of $J$ are reached at about $t=4\,\mathrm{fm\!/\!c}$ independent of the strength of the magnetic field. For the CEOM scenario, the value of $J$ doesn't change with time, as shown in Fig.~1 of Ref.~\cite{Zhou}. As shown in the right panel, the equilibrated value of $J$ increases almost linearly with the strength of the magnetic field, and the values are similar for the SEOM scenario with and without the collision process. It is seen that the equilibrated values from the SEOM are still lower than that from the CEOM with the truncation $0.3<\sqrt{G}<1.7$, with the latter lower than the theoretical limit from Eq.~(\ref{equ:J}) with $\alpha=1$.

    \begin{figure}[ht]
        \includegraphics[angle=0,scale=0.42]{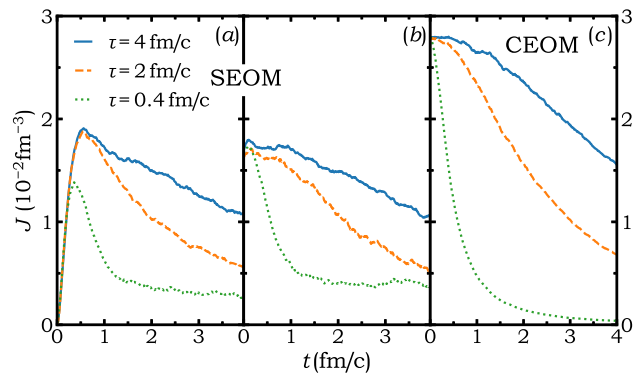}
        \caption{(Color online) Time evolution of the electric charge currents with different damping external magnetic fields from the SEOM initialized with $\vec{\sigma}=c\hat{k}$ (a), the SEOM initialized with an equilibrated distribution (b), and the CEOM (c). }
        \label{fig:cs_J3}
    \end{figure}

    In order to study the situation closer to that in relativistic heavy-ion collisions, we have also done simulations under a damping external magnetic field parameterized as
    \begin{equation}
        eB_y\!\left(t\right) = \frac{eB^0_{y}}{1+\left(t/\tau\right)^2}, \label{equ:mag}
    \end{equation}
    where $eB^0_{y}=0.5$ GeV/fm is the magnetic field at $t=0$ and $\tau$ characterizes its life time. Figure~\ref{fig:cs_J3} displays the time evolution of the electric currents $J$ under different life times of the magnetic field for different scenarios, i.e., with the spin initialized as $\vec{\sigma}=c\hat{k}$ from the SEOM (a), with the initial distribution taken as the equilibrated $\vec{\sigma}$ and $\vec{k}$ distribution at $t=4\,\mathrm{fm\!/\!c}$ under $eB^0_y$ from the SEOM (b), and from the CEOM (c). The collision process is incorporated in all scenarios. It is found that the electric charge current $J$ at later stage is not so sensitive to the initialization for the SEOM scenario, partially due to the collision process that damps oscillations. Although the initial $J$ is larger for the CEOM scenario, it damps more quickly compared with that for the SEOM scenario. This is due to the immediate change of the spin polarization with the magnetic field for the CEOM scenario, while some relaxation process is needed for the SEOM scenario, especially under a rapidly damping magnetic field.


    \begin{figure}[ht]
        \includegraphics[angle=0,scale=0.50]{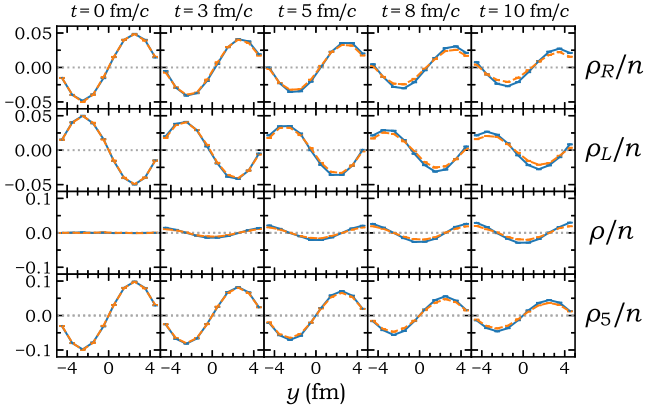}
        \caption{(Color online) Distributions of various reduced densities in $y$ direction at different times characterizing the CMW, with the solid lines from the CEOM and the dashed lines from the SEOM.}
        \label{fig:evo5}
    \end{figure}

\begin{figure}[ht]
        \includegraphics[angle=0,scale=0.50]{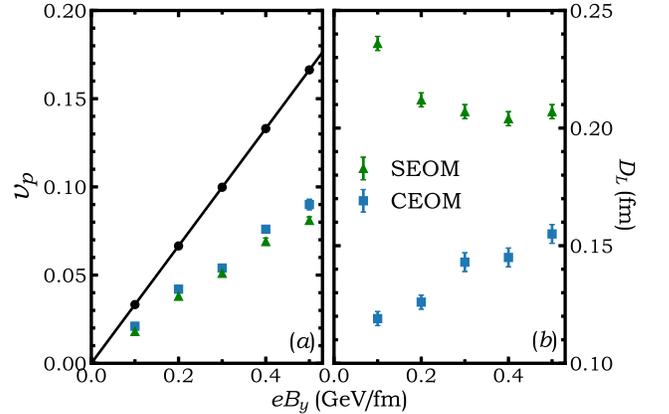}
        \caption{(Color online) Dependence of the phase velocity $v_p$ (a) and the diffusion constant $D_L$ (b) on the strength of the magnetic field from the SEOM and the CEOM, with the solid line representing the theoretical limit.}
        \label{fig:cs_v_p_D_l}
    \end{figure}

    To study the CMW, we set the density distributions in the initial stage as
    \begin{equation}
        \rho_{R\!/\!L}\!\left(y, 0\right) = \pm\frac{1}{2}A_cn\sin{\left(\beta y\right)}, \label{equ:init}
    \end{equation}
    where $\beta=\pi\!/l$ is the wave number, and $A_c=0.1$ represents the maximum chiral asymmetry. Applying the periodic boundary condition $\rho_{R\!/\!L}|_{y=-l}\!=\!\rho_{R\!/\!L}|_{y=+l}$, its expression at time $t$ and position $y$ can be obtained from Eq.~(\ref{equ:cmw}) as
    \begin{equation}
        \rho_{R\!/\!L}(y,t) \approx \pm\frac{1}{2}A_cne^{-D_L{\beta}^2t}\sin{\left[\beta\left(y\mp v_p t \right) \right]}. \label{equ:rhoc}
    \end{equation}
    The electric and axial charge density can be respectively expressed as
    \begin{align}
        \rho &= \rho_R+\rho_L \approx -A_cne^{-D_L{\beta}^2t}\sin{\left(\beta v_p t\right)}\cos{\left(\beta y\right)}, \label{equ:r_charge}\\
        \rho_5 &= \rho_R-\rho_L \approx +A_cne^{-D_L{\beta}^2t}\cos{\left(\beta v_p t\right)}\sin{\left(\beta y\right)}.
    \end{align}
    The time evolutions of distributions of various densities in $y$ direction shown in Fig.~\ref{fig:evo5} are consistent with the above formulas, from the SEOM shown by dashed lines and from the CEOM shown by solid lines, with the collision process incorporated. The phase velocity $v_p$ and the diffusion constant $D_L$ can be obtained by fitting the time evolution of the distribution with the above formulas, and their dependence on the strength of the magnetic field are shown in Fig.~\ref{fig:cs_v_p_D_l}. It is seen that $v_p$ from the CEOM is slightly higher than that from the SEOM due to a weaker CME for the later scenario, while it is still lower than the theoretical limit obtained from Eq.~(\ref{equ:v_p}) with $\alpha=1$. It is noteworthy that the net current is instantaneously induced by the net density from the CEOM, while a relaxation process is needed from the SEOM. The diffusion constant from the SEOM slightly decreases with increasing $eB_y$, and it is larger than that from the CEOM, with the former (latter) decreasing (increasing) slightly with increasing $eB_y$. The decreasing trend of $D_L$ with increasing $eB_y$ from the SEOM is consistent with that observed in Ref.~\cite{Kha111}.

    \begin{figure}[ht]
    	\includegraphics[angle=0,scale=0.55]{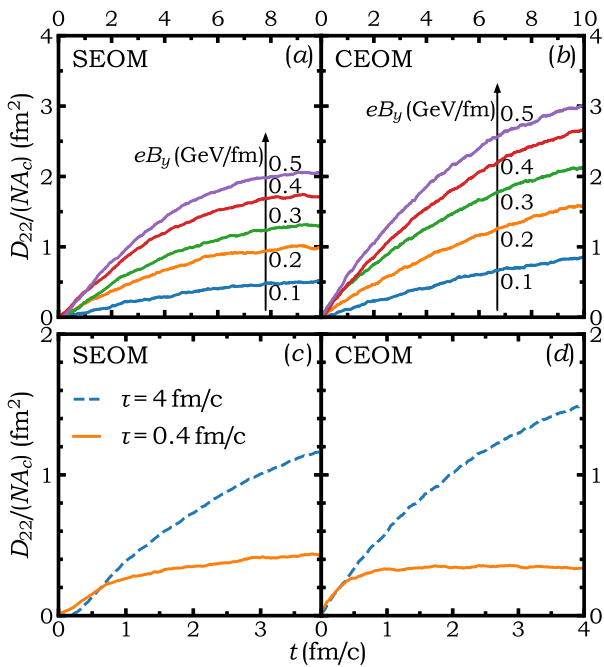}
	    \caption{(Color online) Time evolution of the reduced electric quadrupole moment under different constant (upper) and damping (lower) magnetic fields from the SEOM (left) and the CEOM (right).}
        \label{fig:cs_rupole}
    \end{figure}

    The elliptic flow splitting between particles with oppositive charges originated from the electric quadrupole moment is the key observational consequence of the CMW. To quantify the electric quadrupole moment in the box system, as can be seen from the third row of Fig.~\ref{fig:evo5}, we define the electric quadrupole moment as
    \begin{equation}
    	\mathcal{D}_{22} = \int\rho(\vec{r})\left(3y^2-r^2 \right)\mathrm{d}^3 r. \label{equ:Dint}
    \end{equation}
    Using Eq.~(\ref{equ:r_charge}), the above integral can be carried out as
    \begin{equation}
    	\mathcal{D}_{22} = \frac{4A_cN}{\beta^2}e^{-D_L\beta^2t}\sin(\beta v_p t), \label{equ:D_22}
    \end{equation}
    where $N=8nl^3$ is the total particle number. The upper panels of Fig.~\ref{fig:cs_rupole} display the time evolution of the electric quadrupole moment under different constant magnetic fields. It is seen that $D_{22}$ increases more slowly from the SEOM due to the larger diffusion constant $D_L$ compared with that from the CEOM. The general feature that $D_{22}$ increases with the increasing strength of the magnetic field is observed as expected, but it is of interest to see how is it like in a damping magnetic field. As shown in the lower panels of Fig.~\ref{fig:cs_rupole}, $D_{22}$ increases faster from the CEOM than from the SEOM under the magnetic field with a life time $\tau=4$ fm/c. However, due to the spin relaxation process, the SEOM leads to a larger final electric quadrupole moment than the CEOM under a fast damping magnetic field with a life time $\tau=0.4$ fm/c.


    To summarize, the chiral magnetic effect and the chiral magnetic wave have been studied in a box system with the periodic boundary condition under a uniform external magnetic field based on the spin kinetic equations of motion for massless particles. Although results are qualitatively similar compared with those from the chiral kinetic equations of motion, the spin dynamics leads to weaker chiral effects and a stronger damping of the chiral magnetic wave. Due to the spin relaxation process, however, these chiral effects are less sensitive to the fast damping of the magnetic field. Our study helps to better understand the spin and chiral dynamics in relativistic heavy-ion collisions.

For the chiral kinetic equations of motion, more recent studies have shown that there are additional quantum corrections to the dispersion relation of massless particles under the magnetic field~\cite{CKM3,Man14}. The relation between the more realistic chiral kinetic equations of motion and the spin kinetic equations of motion needs further investigations. For the two-body collision treatment in the present study, it satisfies the energy conservation condition, while the conservation of the angular momentum in each single collision is not guaranteed. In order to satisfy the angular momentum conservation while keeping the Lorentz invariance, one needs to incorporate an additional side jump~\cite{Che14,Che15} in the collision treatment. It is of great interest to investigate chiral effects by doing transport simulations according to the spin kinetic equations of motion or the chiral kinetic equations of motion with quantum corrections by incorporating the side-jump collision treatment in future studies.

We thank Chen Zhong for maintaining the high-quality performance of the computer facility. This work was supported by the Major State Basic Research Development Program (973 Program) of China under Contract No. 2015CB856904 and the National Natural Science Foundation of China under Grant Nos. 11922514 and 11421505.

\end{document}